\documentclass[conference,a4paper]{IEEEtran}

\IEEEoverridecommandlockouts
\usepackage[left=1.59cm, right=1.59cm, top=1.8cm, bottom = 2.75cm]{geometry} %91, 54
\usepackage{cite}
\usepackage{amsmath,amssymb,amsfonts}
\usepackage[ruled, vlined, linesnumbered]{algorithm2e}
\usepackage{graphicx}
\usepackage{textcomp}
\usepackage{xcolor}
\usepackage{eurosym}
\usepackage{dsfont}
\usepackage[tableposition=top]{caption}
\newtheorem{proposition}{\bf Proposition}
\usepackage{url}
\newcommand{\tcr}{}
\def\BibTeX{{\rm B\kern-.05em{\sc i\kern-.025em b}\kern-.08em
    T\kern-.1667em\lower.7ex\hbox{E}\kern-.125emX}}
    
\begin{document}

\title{Impact of Strategic Electric Vehicles Driving Behavior on the Grid
}

\author{\IEEEauthorblockN{Benoît Sohet}
\IEEEauthorblockA{benoit.sohet@edf.fr\\ \textit{LIA, Avignon Univ., France}\\\textit{EDF R\&D, MIRE Dept,}\\ %MIRE
\textit{Paris-Saclay, France}\vspace{-10mm}}
\and

\IEEEauthorblockN{Yezekael Hayel}
\IEEEauthorblockA{yezekael.hayel@univ-avignon.fr\\ \textit{LIA, Avignon Univ., France}\vspace{-10mm}}
\and

\IEEEauthorblockN{Olivier Beaude}
\IEEEauthorblockA{olivier.beaude@edf.fr\\
\textit{EDF R\&D, OSIRIS Dept,}\\ %MIRE
\textit{Paris-Saclay, France}\vspace{-10mm}} %Paris-Saclay
\and

\IEEEauthorblockN{Alban Jeandin}
\IEEEauthorblockA{alban.jeandin@edf.fr\\\textit{EDF R\&D, MIRE Dept,}\\ %MIRE
\textit{Paris-Saclay, France}\vspace{-10mm}}

}
% \vspace{-10mm}

\maketitle

\begin{abstract}
In the context of transport electrification, a model coupling Electric Vehicles (EV) driving and charging decisions is considered. While a Traffic Assignment Problem (TAP) is considered for the driving part, the charging operation is done with an exact load flow on a simple distribution network. This setting allows assessing precisely the coupled impact of driving and charging decisions. In particular, to schedule the charging need coming from the driving part, different charging strategies are defined and compared according to a cost metric based on the load flow computation. In view of this metric, the proposed non-load flow based strategies can perform similarly to a load flow based one, and with significantly reduced computation time and data need. Numerical simulations show how a transportation toll can influence the charging results.%, expressing the need for the coupled analysis of driving and charging decisions. 
\end{abstract}

\begin{IEEEkeywords}
Electric vehicles; Grid interaction
%Congestion game, Electric vehicle, Nonseparable costs, Wardrop equilibrium.
\end{IEEEkeywords}
\vspace{-2mm}
\section{Introduction}
The transport sector is one of the biggest source of gas emissions, with about a quarter of the global level according to the 2017 annual report of the International Energy Agency~\cite{b1}. Full Electric Vehicles (EV) are one of the solutions to the low carbon mobility problematic. However, the increasing number of EV implies a higher power load.
%which, on top of that, is spatially distributed if we think as drivers who charge their vehicles at home.
Then, the behavior of EV drivers, typically the choice of the path to reach their destination and therefore the corresponding energy consumed, induces a demand in electricity from the grid which can be non negligible. Therefore, there is a natural relationship between driving and charging at the user level and globally, between transportation and grid networks.
%Many recent papers are related to this relationships (identified in the recent survey \cite{WEI19}), and our study here is focused on understanding one particular aspect of this complex system. 

The interaction (``coupling") of EV driving and charging decisions is a recent topic of research, as identified in the review paper~\cite{b2}. In~\cite{b3}, only EV are considered, and their charging choices are represented in an ``extended transportation network": each station is replaced by a set of virtual arcs, each one corresponding to a specific charging (energy) amount.
However, there is no smart charging considerations: the resulting aggregated EV charging need is not scheduled in time but instead charged at once.
Moreover, load flow equations considered for the electricity network do not take into account power losses. In~\cite{b4}, Gasoline Vehicles (GV) are added to the problem of~\cite{b3} and more realistic load flow equations are used. Charging need is supposed to be the same for all EV and charging unit prices depend on the total demand at the corresponding charging station.
Still, no algorithm is proposed to schedule the charging operations.
Other works mentioned in~\cite{b2} offer similar coupled models of EV driving and charging but with limited smart charging algorithms.
In~\cite{b5} the vehicle's destination offering a minimal cost is chosen while in~\cite{b6} a fleet operator chooses the proportion of its vehicles to charge instead of taking customers. Finally,~\cite{b7} proposes a planning model where both the transportation and the electricity network are optimally sized, minimizing the operators investment.
The EV charging need comes from a Traffic Assignment Problem model~\cite{b8} and the electricity network is considered, the load flow equations being linearized (DistFlow)~\cite{b9}.

%Many papers focus on a centralized point of view of the transportation network in the sense of autonomous vehicles \cite{} or centralized control of fleet of EV \cite{Sun2019}. This type of models implies large scale optimization techniques and the use of heuristics in order to solve the Traffic Assignment Problem (TAP) \cite{Patriksson}. In our framework, the routing decision is strategically determined by each driver selfishly, and then a non-atomic game and Wardop equilibrium \cite{SHEFFI85} concepts are considered.
%, named a Charging Service Operator (CSO),
Compared to this existing work, one of our previous papers \cite{b10} considers a coupled transportation and charging setting, with both EV and GV, in which a central operator schedules the load of an EV fleet following a water-filling algorithm introduced in~\cite{b11}. % at a particular EV charging station (EVCS).
A typical application is a charging Park \& Ride hub which can be equipped with local electricity production like photovoltaic panels. EV drivers can park there and charge the electricity consumed during the morning while working at a nearby office.
However this previous paper does not study into details the impact on the grid itself of the charging need of EV. %who are charging typically during office hours.  
%is about understanding decentralized routing decisions of EV drivers considering driving congestion from the transportation network, as well as energy cost considering smart charging systems. Particularly, 
In the present paper, the impact of EV charging on the grid is obtained through an exact load flow solution. This load flow is the basis of a charging algorithm, as well as a metric to compare this first algorithm with two heuristic charging policies which are “more distributed”.
The main contributions of this paper are:

\begin{itemize}
    \item the design and numerical comparisons of three different charging scheduling algorithms, depending on which operator manages the charging operation and what information it has access to;
    \item to model and interpret numerically an accurate impact of a transportation incentive (traffic toll) on a medium voltage grid, via the EV coupling of the two systems. This impact is measured at the head of a network, typically a transformer dedicated to a set of Charging Stations (EVCS) with a particular grid topology.
\end{itemize}

The rest of the paper is organized as follows. Section \ref{model} describes the global coupled model incorporating the congestion game and the charging problem. The algorithmic approaches which depend on the managing operator are fully described in section \ref{sec:charg2}. Performances of those approaches are compared and studied using real data sets in section \ref{num}. Finally, main conclusions and perspectives are given in the last section \ref{conc}.

%\newpage

% The charging operation is also taken into account in this work. For simplicity, it is assumed that all Charging Stations (EVCS) are controlled by a unique operator (called Charge Point Operator, or CSO).
% An important feature of this work is smart charging: this CSO schedules the charging operation of all EV inside a long enough time period.
% Therefore, we need to consider situations where vehicles are parked during several hours, such as commuting, where users drive to work and park their vehicle during working hours.
% In this context, EV charge at their destination: 
% therefore, it is assumed that there is an EVCS at each destination.
% EVCS not located at destinations are not used by commuting EV and thus, not considered in this work.
% In other words, EVCS and destinations are equivalent here.

\vspace{-2mm}
\section{Model}
\label{model}

\begin{figure}
    \centering
    \includegraphics[width = 0.4\textwidth]{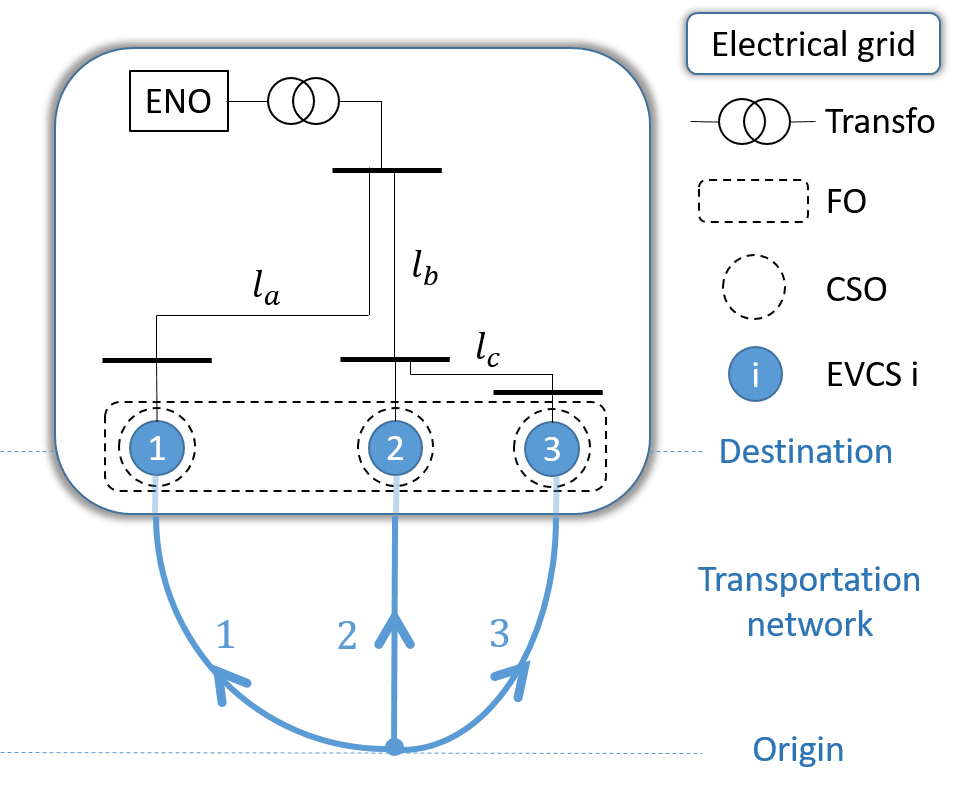}
    \caption{\small Transportation (in blue) and electrical (framed) networks considered in this work.
    ENO, FO and CSO stand for respectively Electrical Network, Flexibility and Charging Service Operators.
    \vspace{-4mm}}
    \label{fig:schema}
\end{figure}
% \begin{figure}
%     \centering
%     \includegraphics[width = 0.4\textwidth]{Images/schema_reseaux.png}
%     \caption{Caption}
%     \label{fig:grid}
% \end{figure}

The use case considered in this work to accurately study the impact of the transportation system on the electrical grid is commuting:
drivers choose one of the three paths of the transportation network of Fig.~\ref{fig:schema} to drive from home to work.
Users choosing path $i$ park their vehicle at the corresponding EVCS $i$, \tcr{from which they can reach the common destination by foot or public transport (which is the purpose of Park \& Ride hubs)}.
%which is assumed to be located nearby the common destination.
These three EVCS belong to the electrical grid of Fig.~\ref{fig:schema} \tcr{and can be up to a dozen kilometers apart}. Then, EV are charged during working hours in a smart manner, chosen by operators willing to minimize their costs.
% Note that these networks were chosen for a better clarity for numerical illustrations, and that the following model can be applied to any network.
This toy example helps to illustrate the importance of a coupled problem of driving and charging, particularly considering the impact on the grid. A more general transportation network and/or electrical grid topology can be considered without changing the main messages and results of our study.

% In this work, vehicles choose a path to reach their destination, rationally and with complete information.
% There can be several classes of vehicles: for example, Electric Vehicles (EV, index \textit{e}) and Gasoline Vehicles (GV, index \textit{g}).
% Note that EV may in turn be divided into several classes, in function of their initial State of Charge (SoC).
% However here, to simplify notations, it is assumed they all share the same SoC, sufficient to complete their trip.

% There, EV can choose their charging amount, provided that their SoC at the end of the day allow them to drive back home, for example.
% This choice is enabled by replacing EVCS (i.e., destinations) nodes by virtual parallel arcs, each one representing the charging quantity desired, as in~\cite{alizadeh2016}. %(see Figure~\ref{fig:schema}).
% For simplicity, GV are not concerned with energy quantity choices through the virtual arcs:
% their energy cost is merely proportional to what they consumed in the morning to get to work.
% Considering the small time window when users choose their path and charging quantity, this problem is modeled through a stationary nonatomic game.
\vspace{-2mm}
\subsection{Congestion game}

The goal of this section is to predict the impact of transportation parameters on the EV charging need at each EVCS.
%The first part of the problem, 
These charging needs depend on how many EV are at each EVCS, so that the path choice made by drivers needs to be modeled.
In this work, this is done via a congestion game: each driver has complete information about the possible costs and is fully rational in the sense that it chooses the path with minimal total cost. This cost depends on the choices of other drivers, in particular on the number of vehicles on each path, through network congestion effects. This is the typical context of congestion game problems \cite{b8}. Two classes of vehicle are considered: EV (associated with subscript \textit{e}) and GV (\textit{g}), which have different costs because of the different energy type. This model works with a larger number of classes, such as different types of EV depending on their initial State of Charge (SoC), but here all vehicles of the same class are assumed to share the same characteristics which correspond to an average use case. 
%Each commuter is associated with either an EV or a GV.
In this transportation network there are $N$ vehicles, with a proportion $X_s$ of vehicles of class $s$ and a proportion $x_{s,i}$ of class $s$ vehicles on path $i$.
Then, for example, the number of EV on path $i$ is $x_{e,i} X_e N$.\\
%Assuming $N$ is large enough, the game can be considered as nonatomic, meaning that a sufficiently small number of vehicles has no effect on the outcome of the game, and that variables $x_{s,i}$ can be treated as continuous.
Vehicles of class $s$ on path $i$ experience three kinds of costs: %(for any $i,s$):

\begin{itemize}

\item Travel duration, given by the following congestion function (from the Bureau of Public Roads~\cite{b12}):
\begin{equation}
d_i(x_i)= d^0_i\left[1+2\left(x_i/C_i\right)^4\right],
\label{eq:BPR}
\end{equation}
where $d_i^{0}=l_i/v_i$ is the minimal travel duration (corresponding to free flow at speed limit $v_i$), with $l_i$ the length of path $i$ and $C_i$ its capacity.
 Note that travel duration is the same for both vehicle classes and depends on the total number of vehicles $x_i=\left(x_{e,i}X_e+x_{g,i}X_g\right)N$ on path $i$.
 \tcr{A stochastic version of travel duration can be readily implemented in our model.
 This paper considers the deterministic function~\eqref{eq:BPR}, which can be seen as the average travel duration.}

\item Energy consumption $l_i m_s \lambda_s$, with $m_s$ the energy (electricity or fuel resp. for $s=e,g$) consumption per distance unit and $\lambda_s$ the energy unit price. Note that $m_s$ is assumed constant and does not depend on speed profiles, thus the energy consumption only depends on traveled distance.

\item Other constant costs, like traffic tolls $t_{s,i}$ imposed by a Transportation Network Operator (TNO).\\
\end{itemize}
\vspace{-4mm}
\noindent
Summing all these costs, the total driving cost for a type $s$ vehicle on path $i$ is given by the following expression:
\begin{eqnarray}
\label{eq:gen_cost}
c_{s,i}(x_i) = \tau d_{i}(x_i)+ l_i m_s \lambda_s+ t_{s,i}\,,
\end{eqnarray}
where $\tau$ is the cost of one unit of time spent driving.

% Each user has complete information about the possible costs and is rational: she chooses the path with minimal total cost, which depends on the choice of others.
By all acting rationally, users will reach a particular distribution of choices between the three paths, denoted by $\mathbf{x}^* = \big(x_{s,i}^*\big)_{s,i}$ and called a Wardrop Equilibrium (WE)~\cite{b13}. %is the italics necessary?
This equilibrium situation gives a model of users behavior in a stable regime where no user has an interest to change her choice unilaterally.
%This is a typical situation after some learning periods when drivers determine their route or follow a guidance app.
The proposed approach can thus be used to evaluate various incentive mechanisms numerically -- in a planning stage or tool -- in order to ``select"  a particular equilibrium before it will be observed in practice (note that this concept is now commonly used in many operational public transportation planning tools for the ``route choice'' step in four-steps models), as~done in Section~\ref{sec:num_sens}.

The total charging need $L_i$ at each EVCS $i$ can then be computed from the WE.
First note that the underlying assumption of the energy consumption term in~\eqref{eq:gen_cost} is that EV want to charge exactly the electricity amount they consumed during their trip.
Then, $L_i$ depends on the proportion $x_{e,i}^*$ of EV parking at EVCS $i$ at WE and on what they consumed to get there, $l_i m_e$:
\vspace{-2mm}
% EV are assumed to charge
% as every EV will ask for charging the amount of energy used in order to reach the EVCS which is related to the path chosen. As expressed in the energy consumption term in~\eqref{eq:gen_cost}, vehicles ask the quantity of energy they have consumed to drive from $O$ to $D$, which is $l_i m_s$. Thus, for each EVCS $i$ the energy demand which corresponds to the charging need for EVs is given by:
\begin{equation}
    L_i = l_i m_e \times x_{e,i}^*X_e N\,.
    \label{eq:Le}
\end{equation}
% \vspace{-2mm}
\tcr{Note that even GV may impact the different charging needs at the EVCSs by adding congestion on some roads which incites EV to favor other roads and thus EVCS.}
% We observe that more EV flow on path $i$, and then at EVCS $i$ induces higher demand in energy at this charging station. 
% \vspace{-10mm}

\vspace{-2mm}
\subsection{Charging problem}
\label{sec:charg1}

Given the charging need $L_i$ at each EVCS $i$ resulting from the commuting game (see~\eqref{eq:Le}), the charging operation of EV during working hours at all EVCS is scheduled by one or several operators (depending on the scenario considered, as explained later in this section). More precisely, the charging period (corresponding to working hours) is divided into several time slots.
Each EVCS $i$ has its own nonflexible consumption $\ell_{i,t}^0$ for each time slot $t$. %local electricity generation $p_{i,t}$ and 
Then, for each EVCS $i$, an operator has to determine the quantity $\ell_{i,t}$ to charge (flexible consumption) at each time slot $t$ in order to minimize its energy costs, and satisfying the total charging need  $L_i$ at this EVCS.
%\begin{equation}
%    \sum_t \ell_{i,t} = L_i (= l_i m_e \times x_{e,i}^*X_e N)\,.
%\label{eq:cstr}
%\end{equation}

Several operators are part of the electrical system, as shown in Fig.~\ref{fig:schema}. Each EVCS $i$ is under the supervision of a Charging Service Operator (CSO). Several CSO may be managed together by what is called a Flexibility Operator (FO). The electrical grid, from the transformer to the EVCS, is managed by an Electric Network Operator (ENO).
%These three types of operators are illustrated in Fig.~\ref{fig:op}.
Depending on which of these three operators controls the charging operation scheduling, three scenarios are considered.
The algorithms solution of these scenarios are detailed in next section.% Let us first describe the three scenarios.
% \begin{figure}
%     \centering
%     \includegraphics[width = 0.4\textwidth]{Images/op.png}
%     \caption{ENO, FO, CSO =}
%     \label{fig:op}
% \end{figure}

\subsubsection{Local (CSO)}
The charging scheduling at each EVCS is done by the corresponding CSO. Each CSO has no knowledge about nonflexible loads and charging profiles chosen by the other CSO, and about the characteristics of the grid. Thus, CSO $i$ minimizes its own energy costs \tcr{(reduced to its contract with the ENO)}, expressed by a quadratic (extendable to other monomials) proxy~\cite{b11} of its total load and satisfying the charging needs:
\begin{equation}
    \min_{\ell_{i,t}} \sum_t \eta_t\left(\ell^{\text{tot}}_{i,t}\right)^2
    \quad \text{s.t.} \sum_t \ell_{i,t} = L_i\,,
    \label{eq:local}
\end{equation}
with $\eta_t$ the weight of time slot $t$, and $\ell^{\text{tot}}_{i,t} = \ell^0_{i,t} + \ell_{i,t}$.%\,. %-p_{i,t}+

%The solution $\left(\ell_{i,t}^d\right)_t$ of this minimization problem (under constraint~\eqref{eq:cstr}) has a water filling structure~\cite{mohsenian10}.
%For a precise solution formulation, please refer to previous works~\cite{SOHET18,sohet2019}.

\subsubsection{Global (FO)}
The charging scheduling of all EVCS is done by the FO.
This aggregator has complete information on all EVCS, but not on the grid.
This way, the FO minimizes the total cost of the EVCS:
\begin{equation}
    \min_{\ell_{i,t}} \sum_t \eta_t\left(\sum_i \ell^{\text{tot}}_{i,t}\right)^2
    \quad \text{s.t.} \sum_t \ell_{i,t} = L_i\,.
    \label{eq:global}
\end{equation}
% with the same constraint equation (\ref{const}) for all $i$ and $t$.
% The associated minimization problem has several optimal charging profiles corresponding to the optimal global charging profile $\left(\ell_{t}^g\right)_{t}$.
% The one chosen here, $\left(\ell_{i,t}^g\right)_{i,t}$, is as close as possible to the distributed profiles $\left(\ell_{i,t}^d\right)_t$ ($\forall i$), i.e., the smoothed total load curve is as smooth as possible locally, at each EVCS:
% \begin{equation}
%     \ell_{i,t}^g = \ell_{i,t}^l + \frac{\ell_{t}^g-\sum_i\ell_{i,t}^l}{T}\,.
% \end{equation}

Note that in the last two scenarios, the operators considered solve optimization problems~\eqref{eq:local} and~\eqref{eq:global} regardless of the grid topology, unlike the ENO in next scenario.
\subsubsection{Grid aware (ENO)}
The charging scheduling of all EVCS is done by the ENO, which has complete information on the EVCS and the grid.
The ENO minimizes the impact of the charging operation on the grid, which is expressed as a monomial of the apparent power $S$ at the head of the grid needed to meet the electricity demand $\ell^{\text{tot}}_{i,t}$ at each EVCS~$i$:
\begin{equation}
    \min_{\ell_{i,t}} \left[\mathcal{G} = \sum_t\eta_t S^2\hspace{-1mm}\left(\ell^{\text{tot}}_{1,t}\,,~ \ell^{\text{tot}}_{2,t}\,,~ \ell^{\text{tot}}_{3,t}\right)\right]
    \text{s.t.} \sum_t \ell_{i,t} = L_i \,.
    \label{eq:ENO}
\end{equation}

%This power $S$ is found by solving an optimal power flow problem using the bus injection model (see~\cite{zhu2015}).
%Unlike in the two previous scenarios, the optimization problem~\eqref{eq:ENO} does not have an explicit optimal profile, but it can be found numerically using convex optimization techniques.\\
Note that in all three scenarios, the various operators all minimize a certain power:
The active powers of each EVCS in the local scenario, the sum of these powers in the global one and the apparent power at the head of the grid for the grid aware one. 

Given the charging needs from previous section induced from the drivers behavior (equation (\ref{eq:Le})), the charging operation is scheduled according to each one of the three scenarios described above. Then, the actual state of the grid corresponding to each scenario loads is computed using the power flow equations introduced in next section. The ultimate goal is to compare the efficiency of each scenario in terms of grid costs defined by function $\mathcal{G}$ (equation \eqref{eq:ENO}) in Section~\ref{sec:num_cmp} considering real data sets.
%(knowing that the centralized scenario is optimal, by definition).
Before that, the algorithms of the three scenarios are detailed in next section.

\section{Algorithms}
\label{sec:charg2}
\subsection{Local scheduling}
Each CSO $i$ can find the charging scheduling solution of the minimization problem~\eqref{eq:local} by using next proposition with $L=L_i$ and $\ell^0_t=\ell^0_{i,t}$: %-p_{i,t}+
\vspace{2mm}
\begin{proposition}
\label{prop:wf}
Given a nonflexible load $(\ell^0_t)_t$, the optimal scheduling $(\ell^*_t)_t$ of the charging need $L$ is:
\begin{equation}
    \ell_t^* = 
    \begin{cases}
    %\frac{L+L^0_{t_0}}{\sum_{s\leq t_0}\frac{\eta_t}{\eta_s}} - \ell^0_t\,, % t_0
    \left(L+L^0_{t_0}\right)/\left(\sum_{s\leq t_0}\frac{\eta_t}{\eta_s}\right) - \ell^0_t\,, &\text{if } t\leq t_0\,,\\
    0 &\text{if } t> t_0\,,
    \end{cases}
    \label{eq:wf}
\end{equation}
with $L^0_{t} = \sum_{s\leq t}\ell^0_s$, where $t_0$ is such that $L\in ]L_{t_0};L_{t_0+1}]$ and $L_t = \left(\sum_{s\leq t} \eta_{t}/\eta_s\right)\ell^0_t-L^0_t$. %\left(\ell^0_t -\ell^0_s\right) % = (t-1)\ell^0_t - L^0_{t_0 -1}
\end{proposition}
\vspace{2mm}
The optimal charging profile given in proposition~\ref{prop:wf} has a ``water filling'' structure~\cite{b11} when $\eta_t = \eta$.
%Note that this algorithm can be adapted to more realistic scheduling problems, such as cost $G_i = \sum_t \eta_t \ell^2$ time dependant and power constraints $\ell_t \leq \bar{\ell_t}$.
For more details, please refer to our previous work~\cite{b14}.

\subsection{Global scheduling}
Applying proposition~\ref{prop:wf} to all EVCS aggregated (i.e., $L=\sum_i L_i$ and $\ell^0_t=\sum_i \ell^0_{i,t}$) gives an optimal aggregated profile $\left(\ell^*_t\right)_t$ which minimizes $\sum_t \eta_t\ell_t^2$. %-p_{i,t}+
However, several total profiles $\left(\ell_{i,t}\right)_{i,t}$ verify $\sum_i \ell_{i,t} = \ell^*_t$ ($\forall t$), i.e., are solution of~\eqref{eq:global}.
The ``disaggregation" ($\left(\ell^*_t\right)_t \rightarrow \left(\ell_{i,t}\right)_{i,t}$) chosen is presented in Algorithm~\ref{algo:reinforcement}:

%       $\ell_{i,s} \longleftarrow \ell_{j,s} + \frac{1}{(T-1)(N-1)} \ell_{i,t}$\;
\begin{algorithm}%[H]
\KwIn{$\ell_{i,t}^0\,, L_i$} %-p_{i,t}+
$\ell_{i,t} \longleftarrow $ Optimal local profile\;
$\ell_t \longleftarrow$ Optimal aggregated profile\;
$\ell_{i,t} \longleftarrow \ell_{i,t} + \frac{1}{T}\left(\ell_t-\sum_i\ell_{i,t}\right)$\;
    \While{$\exists~ \ell_{i,t}<0$}
    {
    $\ell_{j,t} \longleftarrow \ell_{j,t} + \frac{1}{N-1}\ell_{i,t}$ \hfill $j\neq i$\;
    $\ell_{i,s} \longleftarrow \ell_{j,s} + \frac{1}{T-1} \ell_{i,t}$\hfill $~s\neq t$\;
    $\ell_{j,s} \longleftarrow \ell_{j,s} - \frac{1}{(T-1)(N-1)} \ell_{i,t}$\hfill $j\neq i$, $s\neq t$\;
    $\ell_{i,t} \longleftarrow 0$\;
    }
\KwOut{$\ell_{i,t}$}
\caption{Global scheduling}
\label{algo:reinforcement}
\end{algorithm}

To obtain this optimal global scheduling, the solution of~\eqref{eq:local} (Line 1) is modified as little as possible in order to be solution of~\eqref{eq:global} (Line 3).
Unfortunately, these perturbations create negative charging quantities (Line 4), which are taken into account (Line 8) and compensated (Lines 5, 6 and 7) so that the constraints at each EVCS are still verified.
\subsection{Grid aware scheduling}
The objective function $\mathcal{G}$ of the ENO requires the apparent power $S$ needed to meet the electricity demand at all EVCS
(equation~\eqref{eq:ENO}).
This quantity is obtained by solving the power flow equations from the Bus Injection Model \tcr{(\textit{runpp} function in \textit{pandapower} Python library)} and which corresponds to the power balance at each bus (between the given power production/load $S_{0,k}$ at bus $k$ and power flows $S_k$ from/to the bus):
\vspace{-2mm}
\begin{equation}
S_{0,k} = U_k \sum_{m\in X_k} \overline{Y_{k,m}}\overline{U_m} ~(= S_k)\,,
\label{eq:pf}
\end{equation}
with $U_k$ the complex voltage at bus $k$, $X_k$ the set of buses connected to bus $k$ and $Y_{k,m}$ the admittance of the line between buses $k$ and $m$.

Because of the implicit nature of $\mathcal{G}$, iterative water filling algorithms applied to problem~\eqref{eq:ENO} do not result in an explicit solution, as in the local and global scenarios.
% A natural algorithm to minimize $\mathcal{G}$ is an iterative water filling one:
% Increment $\ell_{i,t}$ associated with the lowest marginal cost $\partial\mathcal{G}/\partial\ell_{i,t}$ and so on, until all EV are charged.
% Unfortunately, as all marginal costs change at each step, the increment step would have to be extremely small.
Instead,~\eqref{eq:ENO} is seen as a convex optimization problem, solved by built-in \tcr{Python function \textit{minimize}, relying on a sequential least squares programming method.}

\vspace{-2mm}
\section{Numerical illustrations}
\label{num}
\subsection{Impact of transportation system on the grid}
\label{sec:num_sens}
\begin{figure}
    \centering
    \includegraphics[width = 0.47 \textwidth]{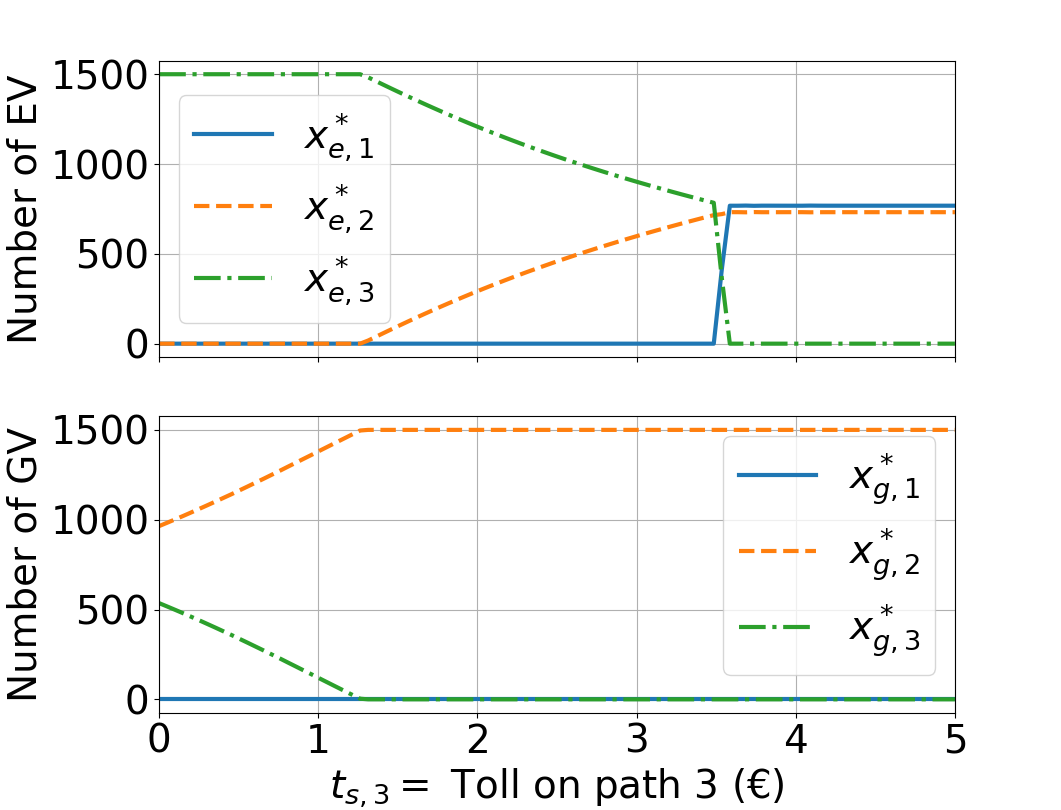}
    \caption{\small Number of EV and GV at equilibrium on each path, in function of toll $t_{s,3}$ on path 3.\vspace{-4mm}}
    \label{fig:we}
\end{figure}

This section illustrates the impact of the transportation system on the electrical grid.
More precisely, we look at how a transportation parameter -- here, the toll on path 3 -- controlled by a Transportation Network Operator (TNO), can impact the grid costs $\mathcal{G}$ related to the charging operation of EV and defined in~\eqref{eq:ENO}.
Note that the toll on path 3 is the same for EV and GV.

The parameters of the congestion game are set as follows.%, unless otherwise specified
The length of the three transportation paths are $l_1 = 30$ km, \tcr{the average commuting driving distance in USA}\footnote{\tcr{\textit{Omnibus Household Survey}, Bureau of Transportation Statistics}}, and $l_2 = l_3 = 20$ km.
The corresponding speed limits are $v_1 = v_2 = 50$ km/h and $v_3 = 70$ km/h, so that $d^0_3<d^0_2<d^0_1$.
The path capacities are equal to the total number of vehicles: $C_i=N = 3000$ vehicles.
The values of the following parameters are the same as in~\cite{b10}: $X_e=50$\%, $\tau=10$ \euro/h, $m_g=0.06$ L/km, $\lambda_g=1.5$ \euro/L and $m_e=0.2$ kWh/km.
The charging unit price is set to $\lambda_e=20$ c\euro/kWh.
There are no tolls on paths 1 and 2: $t_{s,1} = t_{s,2} =0$ \euro.
% The penetration of EV is $X_e=50$\% (and $X_g=1-X_e=0.5$).
% The value of time is set to $\tau =10$ \euro/h, according to a French government report\footnote{\url{http://www.strategie.gouv.fr/sites/strategie.gouv.fr/files/archives/Valeur-du-temps.pdf}.}.

Under these driving conditions, the paths choice of commuters (i.e., the Wardrop Equilibrium $\mathbf{x^*}$) was computed for each toll value $t_{s,3}$ on path 3  (see Fig.~\ref{fig:we}).
When there is no toll (i.e. $t_{s,3}=0$), most vehicles choose path 3 as it is the fastest, except for a few GV which use path 2, the second fastest (otherwise, path 3 would be too congested).
For $t_{s,3}\leq 3.50$ \euro, more vehicles rather choose path 2 because of the toll on path 3.
For $t_{s,3}>3.50$ \euro, path 3 has become even less attractive than the longer path 1, so that no vehicles are left on path 3.

\begin{figure}
    \centering
    \includegraphics[width = 0.5 \textwidth]{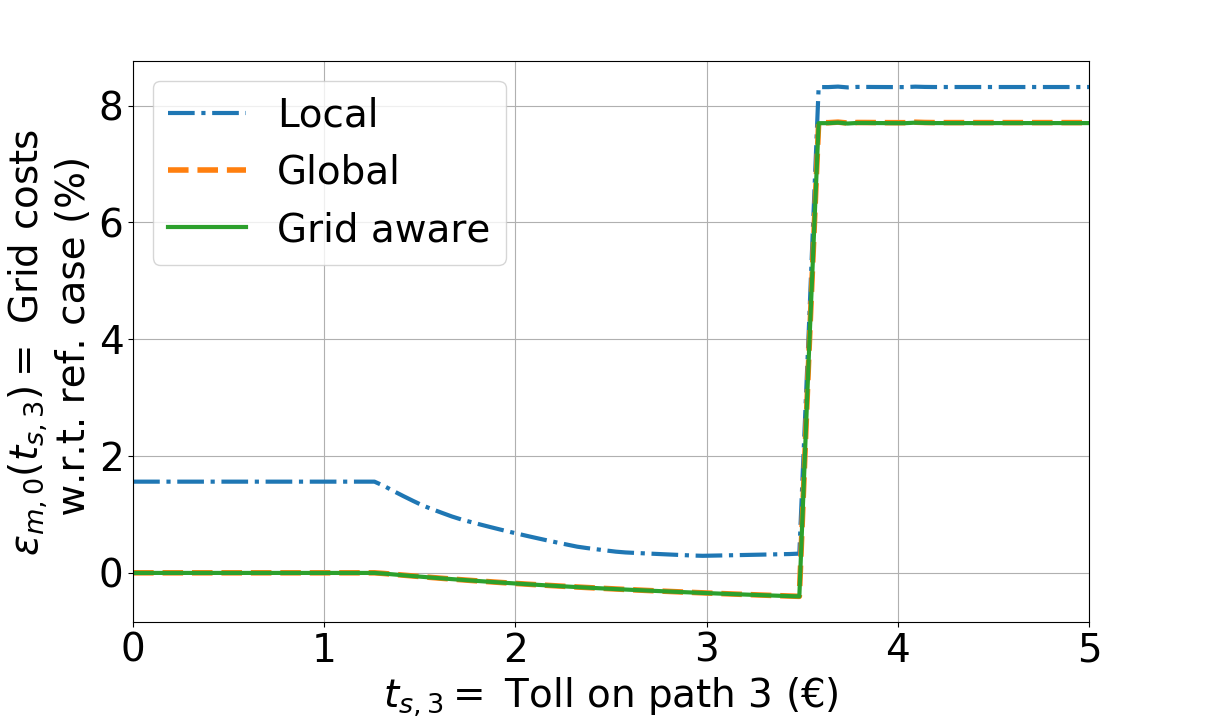}
    \caption{\small Grid costs $\mathcal{G}_m(t_{s,3})$ relative to the reference case (grid aware method with no toll), in function of the toll $t_{s,3}$ on path 3, for each algorithm $m$.
    \vspace{-4mm}}
    \label{fig:impact}
\end{figure}

These path choices made by EV determine the charging need at each EVCS in function of $t_{s,3}$.
Fig.~\ref{fig:impact} then shows the resulting grid costs $\mathcal{G}_m(t_{s,3})$ defined in equation~\eqref{eq:ENO} for each scheduling method $m$ ($m=l,g,a$ resp. for local, global and grid aware), in function of $t_{s,3}$. These costs are normalized into $\varepsilon_{m,0}$, using the grid costs of the grid aware method with $t_{s,3}=0$:
\vspace{-2mm}
\begin{equation}
    \varepsilon_{m,t}(t_{s,3}) = \frac{\mathcal{G}_m(t_{s,3})-\mathcal{G}_a(t)}{\mathcal{G}_a(t)}\,.
\end{equation}

The parameters of the scheduling problem are set as follows\footnote{\url{https://pandapower.readthedocs.io/en/v2.2.0/std\_types.html}.}:
Standard types were used for the transformer (63 MVA 110/20 kV) and the lines (1x240 RM/25 12/20 kV).
The lengths of these lines are $l_a=10$ km and $l_b=l_c=5$ km.
The number of time slots for the scheduling is $T = 8$ and the different objectives defined in~\eqref{eq:local},~\eqref{eq:global} and~\eqref{eq:ENO} are assumed to be time-independent ($\eta_t = \eta$).
Here, the nonflexible load of each EVCS is taken proportional to the consumption of a Texan household (from 9 a.m. to 5 p.m., i.e., eight hours), so that the global demand over the working hours is $\sum_i \sum_{t=1}^T \ell^0_{i,t} = 30$ MWh.
The data set used gives hourly electric consumption throughout a year of Texan households\footnote{Data available at \url{http://www.pecanstreet.org/}.}.
%Toll on path 3 was fixed to $t_{s,3}=4$ \euro, so that there is no charging need at EVCS 3.

The first insight given by Fig.~\ref{fig:impact} is that a transportation toll (between 1.50 \euro~ and 3.50 \euro) can be beneficial in terms of grid costs. %(reduction of the order of 1\%).
For such values of toll, some charging need of EVCS 3 is shifted to EVCS 2 (see Fig.~\ref{fig:we}), which is less costly because EVCS 3 is at the end of the line supplying the two EVCS.
The transportation network impacts directly the grid when EV drivers switch (at $t_{s,3} = 3.50$ \euro) from path 3 to path 1, which is associated with a larger electricity consumption (because longer distance traveled):
This larger charging need causes a grid costs increase of close to 8\%.
Finally, note that the local method gives grid costs only 1\% higher (approximately) than the other two methods. %, which seems to have an important efficiency.
The difference of grid costs between the three methods will be studied in more details in next section.

\subsection{Grid costs associated with scheduling methods}
\label{sec:num_cmp}

\begin{figure}
    \centering
    \includegraphics[width = 0.47 \textwidth]{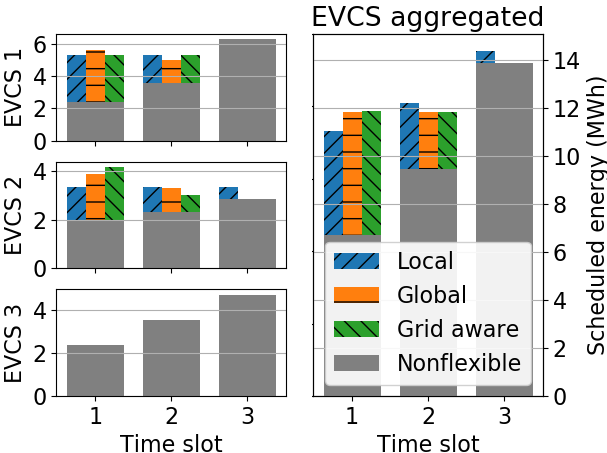}
    \caption{\small Example of the three scheduling algorithms.
    \vspace{-4mm}}
    \label{fig:profile}
\end{figure}

The three scheduling methods introduced in sections~\ref{sec:charg1} and~\ref{sec:charg2} are illustrated in this section, and compared with respect to their associated grid costs. First, an example of these three methods is given in Fig.~\ref{fig:profile}.
The nonflexible consumption (in grey) is the same as in previous section, except that here the eight working hours are split into $T=3$ time slots.
The toll on path 3 is fixed to $t_{s,3} = 4$ \euro, so that there is no charging need at EVCS 3.

The local method smooths each EVCS $i$ load profile $\left(\ell^{\text{tot}}_{i,t}\right)_{1\leq t \leq T}$ (upward diagonal hatch in Fig.~\ref{fig:profile}).
Unfortunately, in some scenarios like here, the corresponding aggregated load profile can be far from smoothed (right figure).
This is the reason why the local method results in higher grid costs:
For example, charging vehicles at EVCS 2 during time slot $t=3$ while there is already a high nonflexible consumption (possibly at other EVCS) can be costly.
The global method smooths the aggregated load profile of the three EVCS (horizontal hatch), with local profiles at each EVCS as smoothed as possible (see left figure).
The optimal ``Grid aware" method, minimizing grid costs, has almost smoothed aggregated and EVCS 1 profiles:
As EVCS 1 is farther away from the transformer than EVCS 2 (10 km instead of 5), charging there is more expensive for the grid.

Then, a comparison between the three methods is given in Table~\ref{tab:perf}.
For this, a thousand nonflexible profiles with $T=8$ time slots are randomly generated for each EVCS, such that for each generation, $\sum_i \sum_{t=1}^T \ell^0_{i,t} = 30$ MWh.
Each generated profile has also a 2 and a 4 time slots version.
This table shows the mean over these random profiles of the normalized grid costs $\varepsilon_m=\varepsilon_{m,4}(4)$ and the execution time $T_m$ of method $m$.
This table confirms that the grid aware method is optimal with respect to grid costs ($\varepsilon_l, \varepsilon_g>0$) and shows that the global one remains very close, while the local method difference is of the order of the percent.
In terms of execution times obtained with an Intel Core i7-6820HQ 2.70GHz, the local and global methods are negligible compared to the grid aware method.
%More than a second per execution.

The execution time depends on the number of variables of the optimization problems~\eqref{eq:local},~\eqref{eq:global} and~\eqref{eq:ENO}: the number of EVCS multiplied by the number of time slots.
A comparison between 2, 4 and 8 time slots shows that the execution time of the grid aware method goes from 1s to approximately 21s.
However, the normalized grid costs of the other two methods increase too (nearly proportional to the number of time slots).
Thus, the choice of the method is a trade-off between execution time and optimal grid costs.
Nevertheless, the global method seems to be the best choice as it is fast and near-optimal, even if it does not take into account the network topology.
Though note that for more complex electrical networks, this might not be the case anymore.

\begin{table}%[h]
\centering
    \caption{\small Depending on the number of time slots into which working hours are divided, mean over 1000 randomly generated nonflexible profiles, of normalized grid costs $\varepsilon_m=\varepsilon_{m,4}(4)$ and execution time $T_m$ of method $m$.}%} Depending on the number of time slots into which working hours are divided, mean over 1000 randomly generated nonflexible profiles, of normalized grid costs $\varepsilon_m=\varepsilon_{m,4}(4)$ and execution time $T_m$ of method $m$
\begin{tabular}{|c||c|c||c|c|c|}
       \hline
Nb time slots & $\varepsilon_l$ (\%) & $\varepsilon_g$ (\%) & $T_l$ (s)& $T_g$ (s) & $T_a$ (s)\\
       \hline
       \hline
2 & 0.4 & 8e-03& 5e-05 & 9e-05 & 0.8 \\ 
       \hline
4 & 1.0 & 2e-02 & 6e-05 & 3e-04 &  4.3\\
       \hline
8 & 2.2 & 3e-02 & 6e-05 &  1e-03 & 20.9
       \\ \hline
   \end{tabular}
    \label{tab:perf}
    \vspace{-3mm}
\end{table}

\section{Conclusion}
\label{conc}

In this paper, the driving and charging decisions of Electric Vehicles (EV) are modeled through a coupled Traffic Assignment Problem and an exact load flow on a simple distribution network. The EV load scheduling is done with three charging strategies, with different assumptions regarding the data needed: from a purely per station distributed strategy, to a centralized one based on the load flow calculation. Numerically, the impact of a transportation parameter - a toll value - is observed on the grid cost obtained on the charging operation; it shows both interlinked driving and charging effects. It also exhibits that distributed charging strategies, which are non-load flow based, perform comparably to the load flow based strategy in terms of grid cost. This opens the way to the use of distributed simpler strategies in this coupled context; however such comparative study will be extended to more complicated electricity networks where this conclusion may differ.
\tcr{In a future work, dynamic charging pricing to make EV ``grid-aware" (by charging at less congested EVCS for example) will be integrated and the different operators and their interactions will be considered.}
%This is part of our future work about this model. 

% \bibliographystyle{ieeetr}
% \bibliography{myrefs.bib}

\end{document}